Research Article

# Maxwell Displacement Current and Nature of Jonsher's "Universal" Dynamic Response in Nanoionics

Alexandr Despotuli, and Alexandra Andreeva

Institute of Microelectronics Technology and High Purity Materials, Russian Academy of Science, 142432 Chernogolovka, Moscow Region, Russia
despot@ipmt-hpm.ac.ru

**Abstract**

New notion - Maxwell displacement current on potential barrier - is introduced in the structure-dynamic approach of nanoionics for description of collective phenomenon: coupled ion-transport and dielectric-polarization processes occurring during ionic space charge formation and relaxation in non-uniform potential relief. We simulate the processes: (i) in an electronic conductor/advanced superionic conductor (AdSIC) ideally polarizable coherent heterojunction, (ii) in a few strained monolayers of solid electrolyte (SE) located between two AdSICs forming coherent interfaces with SE. We prove the sum of ionic current over any barrier and Maxwell displacement current through the same barrier is equal to the current of current generator. "Universal" dynamic response, $\mathrm{Re}\sigma^*(\omega) \propto \omega^n$ ($n < 1$), was found for frequency-dependent conductivity $\sigma^*(\omega)$ for case (ii) with an exponential distribution of potential barrier heights in SE. The nature of phenomenon is revealed. The amplitudes of non-equilibrium ion concentrations (and induced voltages) in space charge region of SE change approximately as $\propto \omega^{-1}$. Amplitudes yield a main linear contribution to $\mathrm{Re}\sigma^*(\omega)$. The deviation from linearity is provided by the cosine of phase shift $\varphi$ between current and voltage in SE-space charge but $\cos\varphi$ depends relatively slightly on $\omega$ (near constant loss effect) for coupled ion-transport and dielectric-polarization processes.

## 1. Introduction

The canonical physical-mathematical formalism for the description of ion transport in solids [1] is based on the concept of a regular crystalline potential relief, i.e. the heights $\eta$ of potential barriers are *const*. As a consequence, ion-transport characteristics (mean-square displacement, diffusion coefficient, mobility, activation energy) have a clear physical meaning only at averaging in the scale much exceeding the length of an elementary ion jump $\Delta$. However, the variation of $\eta$ with a space coordinate $x$ can be considerable in objects of solid state ionics. For example, local regions with $\eta$ variation about 1 eV/nm exist in disordered solid electrolytes (SE) [2-4] and in crystals of superionic conductors (SIC) [5]. Various degrees of ordering occur near surfaces and in the region of interphase and intercrystalline boundaries [6] where specific ion dynamics can arise on the nanoscale [7] due to a non-uniform potential relief. The interface structure exhibits itself in, e.g., frequency-capacitance characteristics of EC/SIC heterojunctions [8-10]. The classification of solid ionic conductors [8-10] distinguishes a class of advanced superionic conductors (AdSIC) among SE and SIC because AdSIC-crystal structure is close to optimal for fast ion transport (FIT). This crystal structure determines the record high level of ion-transport characteristics. The FIT structure represents a rigid sub-lattice of immobile ions in which there are the crystallographic positions interconnected with each other by low potential barriers (conduction tunnels) for hopping mobile ions. The number of such positions is several times larger than number of mobile ions. The network of crystallographic tunnels with potential barriers about 5 $kT$ exists in AdSICs.

In physical chemistry and electrochemistry, much attention is given to static properties of equilibrium distribution of ionic space charge in an electric double layer (EDL) [11-16]. A fewer number of works is devoted to the EDL dynamics on a blocking electrode especially under conditions of steric effects, i.e. a change of mobile ion concentrations near the electrode is limited by the number of available sites in an electrolyte [14,15,17]. In work [18] a frequency-dependent polarization length $\Lambda(\omega)$ as well as two characteristic frequencies corresponding to the charge relaxation in a dense and diffusive layers of EDL were introduced to consider the dynamic properties (at impedance boundary conditions) in terms of the Debye screening length. In [19] the attention is drawn to the fact that the Maxwell displacement current should be taken into account for the process of EDL charging. However, this fundamental current has not been even mentioned in the reviews [2-4] devoted to multiple-aspect discussions on dynamic properties of solid ionic conductors.

Recently, a structure-dynamic approach [20-25] has been proposed for the development of nanoionics [26], particularly, for detailed description of processes in the region of EC/AdSIC coherent ideally polarizable heterojunctions. The approach includes a structure-dynamic model of a nanostructure, a method of "hidden" variables and a physico-mathematical formalism which operates these variables. The model considers ion-transport processes in the EDL region as the movement of mobile ions in the crystal potential relief of a "rigid" sublattice distorted on the boundary. The model describes ion transport in terms of mobile ion concentrations $n_j$ on the crystallographic planes $X^j$ of an ionic conductor in the region of non-uniform potential relief. The $X^j$ planes are situated perpendicularly to direction of an electric field and correspond to the minima of potential relief. At present, non-equilibrium values of these concentrations cannot be determined experimentally, therefore we call them "hidden" variables. The physico-mathematical formalism is based on the principle of a detailed balance and the kinetic equation in the form of the particle conservation law. Computer experiments showed [21-25] that energy dissipation on potential barriers in the fine structure of a space charge on EC/SE heterojunctions decreases considerably with increase of the frequency $\omega$ of current generator. So, the character of a heterojunction response to an external ac influence changes from resistive (ionic currents over barriers) to dielectric, i.e. local Maxwell displacement currents (the quantity directly proportional to the rate changing of electric induction) should be taken into direct account in nanoionics.

This work continues to develop the structure-dynamic approach [20-25] which is used here to model processes (i) in the region of space charge on an ideally polarizable (blocking) heterojunction $EC_b$/AdSIC, and (ii) on a model nanostructure, representing a few strained monolayers (with total thickness $L$) of SE placed between two AdSICs forming coherent interfaces with SE. To determine new elements necessary for the development of a generalized FIT-theory on nanoscale, we introduce a new concept – the Maxwell displacement current on a potential barrier - to describe local ion-transport and polarization processes during space charge formation and relaxation in a non-uniform potential relief.

## 2. The model of a non-uniform potential relief for coherent heterojunction

The theory of dynamic response of an ionic conductor (with high variation of $\eta$ value on the nanoscale) is not developed still. In this section we briefly formulate a theoretical 1-D hopping model of AdSICs with FIT-tunnels and motivate the introduction of the local Maxwell displacement currents into the structure dynamic approach of nanoionics.

The coherent (structure-ordered) $EC_b$/SE-AdSIC ideally polarizable heterojunction is marked in Fig.1A, as the $EC_b/\{X^j\}$ ($j =1,2…M$) where $\{X^j\}$ is the sequence of crystallographic $X^j$ planes, parallel to the $EC_b$-blocking electrode. The $X^j$ planes with $x_j$ coordinates correspond to local minima of non-uniform potential relief in AdSIC. The normal to the electrode surface plane ($X^0$) coincides with the orientation of FIT- tunnels in AdSICs [27] and with the direction of an electric field in EDL. External alternating current influence on heterojunction generates a non-equilibrium ionic space charge in a non-uniform potential relief. At the relaxation, this space charge is transformed into an ionic plate of EDL. Structural coherence of $EC_b$/AdSIC means that the mismatch elastic deformations do not cause essential changes in structure of the rigid sublattice of AdSIC which defines the 1-D FIT-tunnels. However, the mismatch deformations (less 3-5%) can cause the significant changes in a potential relief inside tunnels [28].

The proposed structural model of a heterojunction partitions the 3D space of an ionic conductor (adjacent to $EC_b$) into a number of planar layers (layers between neighboring $X^j$ and $X^{j+1}$ planes), just as the model [11-13] of layered lattice gas in the equilibrium EDL theory. Unlike [11-13], the $X^j$ planes in our model also correspond to the minima of a potential relief for mobile ions [20,25].

For model coherent heterojunction $EC_b$/AdSIC (Fig.1) the number of layers equals $M$. The non-uniform potential relief $\eta(x)$ extends from $EC_b$ along the $x$-coordinate for several nanometers, which is determined by the influence of interface atomic structure. The coordinate $x_0 = 0$ corresponds to the surface of $EC_b$ blocking electrode and $x = x_M$ corresponds to the right edge of a non-uniform potential relief. The symbol $\eta_{j,j+1}$ denotes the height of a potential barrier between neighboring minima "$j$" and "$j+1$". The barrier minima $\eta(x_j) = 0$ are located in the points $x_j$ ($j = 1,2…M$), and the maxima are in ($x_j - \Delta/2$) along the $x$-coordinate, where $\Delta = x_{j+1} - x_j$ . The $\eta$ value decreases with increasing "$j$", i.e. with the distance from $x_0$. The $\eta$ value $\eta$ becomes equal to the barrier height $\eta_v$ in the AdSIC bulk (at $x = x_M - \Delta/2$). Along the $y$ and $z$-coordinates the quantity $\eta(x_j - \Delta/2, y_p, z_q)$, like along the $x$-coordinate, is a discrete function ($p$ and $q$ are integer numbers from 1 to $\infty$) determined by the AdSIC periodicity.

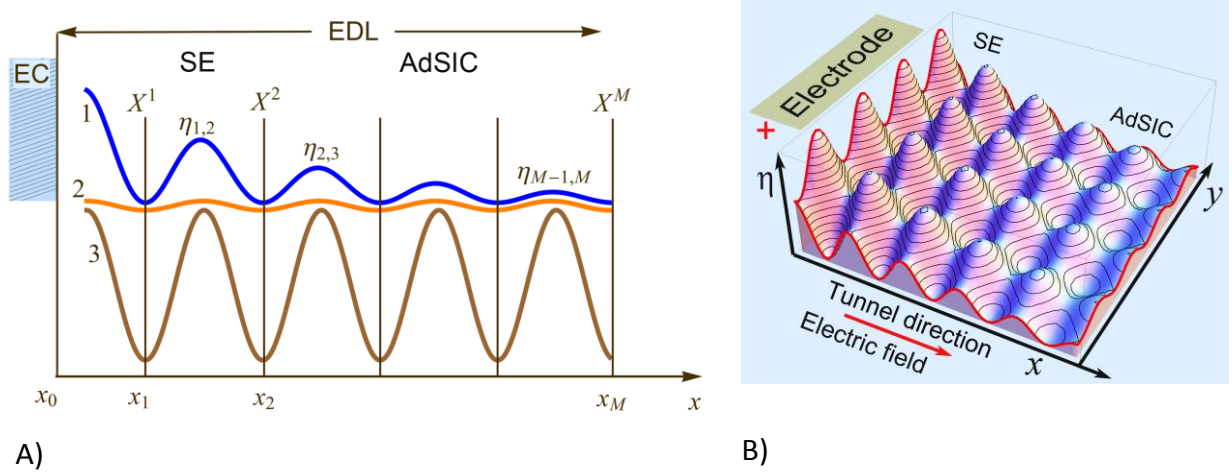

A)  B)

**Figure 1**: The models (A and B) of a non-uniform potential relief with conduction tunnels (a set of low potential barriers). The region of electric double layer (EDL) with large heights of potential barriers is designated as SE. For comparison, the model A) shows the potential reliefs in the region of EC/SE-AdSIC heterojunction (1), in the volume of AdSIC (2) and in a classical ionic crystal (3).

The values of $\eta(x_j - \Delta/2, y_p, z_q) = \eta(x_j - \Delta/2)$ for any $p$ and $q$, i.e. layers between planes with coordinates $x_j$ and $x_{j+1}$ are considered as macroscopic along the $y$- and $z$-coordinates Therefore in the Cartesian $xyz$-system the state of mobile cation is uniquely determined by the $x_j$ coordinate of the $X^j$ plane. Such presentation was chosen because in the rigid sub-lattice of AdSIC there are FIT-"tunnels" of definite crystallographic direction with low potential barriers. 1-D approximation catches the main peculiarities of ion transport in ionic conductors with conduction tunnels.

Figure 2 graphically presents the movement of mobile silver cations through vacant tetrahedral positions in a rigid sub-lattice formed by iodine anions in the $\alpha$-AgI AdSIC. The number of vacant positions available for Ag+ cations is 6 times larger than the number of cations. The projection of a "zig-zag" movement of Ag+ in FIT-tunnels coincides with the electric field direction.

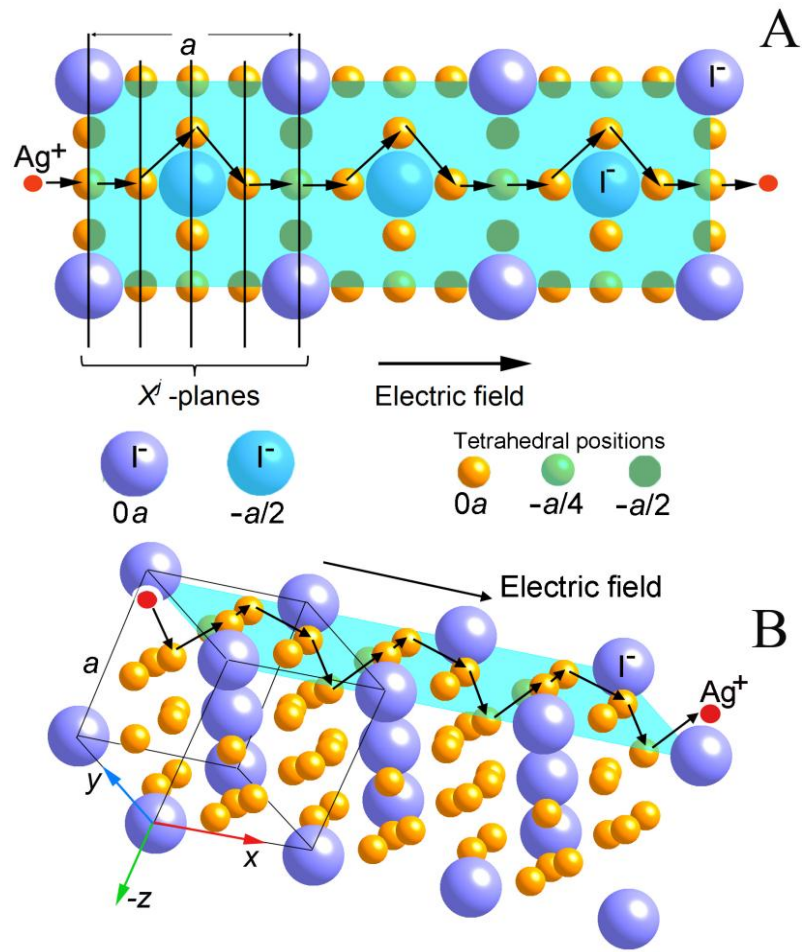

**Figure 2**: Crystallographic structure of the $\alpha$-AgI AdSIC (Im3m, $a \approx 0.509$ nm) in two image projections: A) and B). The "rigid sub-lattice of AdSIC is composed by large iodine anions (blue spheres). Within the I⁻-anion sub-lattice there is a set of interconnected tetrahedral crystallographic positions (small orange spheres), forming the <100> FIT- tunnels through which the Ag⁺ cations (small red circle) can drift in the direction of electric field. For the A) image projection, different colors show the distance of anions I⁻ and tetrahedral positions available for mobile Ag⁺ cations from the crystallographic plane (001).

The proposed structural model simplifies a problem of presentation of real crystal structure with FIT- tunnels. We suppose that the chemical potential of cations does not depend on the coordinate $x_j$, without external influence of current generator on the $X^0$ and $X^M$ edge planes of heterojunction. Therefore the initial equilibrium concentrations of cations equal $n_0$ on all $X^j$ ($j = 1,2…M$) planes. The current generator specifies a current density on the $X^0$ and $X^M$ edge planes of model nanostructure that leads to changes of concentrations of mobile cations on all $X^j$ planes due to long-range electric field. In a non-uniform potential relief the changes of concentrations $n_j$ are different, i.e. a non-equilibrium ionic space charge appears. The appropriate ion-transport and dielectric-polarization processes can be described by the system of differential equations. A detailed description of basic elements of the structure-dynamic approach in nanoionics is given in [25]. By parameter selection, the using system of differential equation [25] can be approximated to the case of different cation concentrations on the $X^j$ planes.

## 3. Objects and conditions of computer experiments

Let consider the objects and conditions of computer experiments.

1. Crystallographically narrow coherent $EC_b$/SE-AdSIC heterojunctions with the variation of $\eta$ value ~0.5eV/nm for which stresses and corresponding deformations spread from the boundary to a distance not longer than several lattice parameters [29,30].

The idea of using coherent $EC_b$/SE-AdSIC heterojunctions for the creation of micron-size supercapacitors with high frequency-capacitance characteristics (nanoionics supercapacitors) was put forward in 2003 [31]. The change in potential barrier heights at the $EC_b$/SE-AdSIC in the EDL region was taken equal to $0.5 \div 0.1$ eV (Fig.3), which corresponds to the experimental data in [10], where the prototypes of nanoionics supercapacitors on the basis of AdSIC were created and investigated.

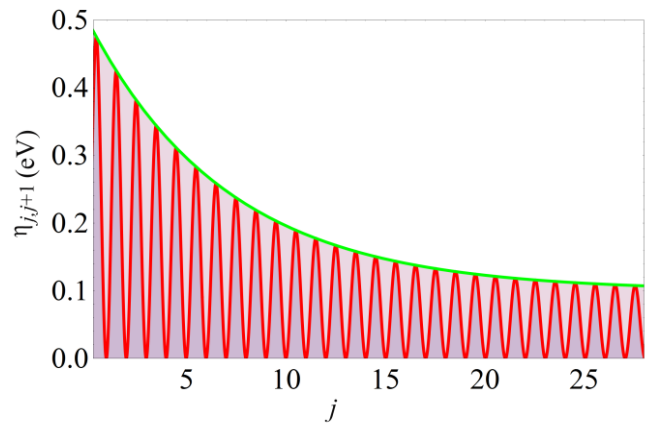

**Figure 3**: The profile of a potential relief (the distribution of barrier heights $\eta_{j,j+1}$ as a function of index $j$) at the modeling of the processes in the region of an $EC_b$/SE-AdSIC coherent blocking heterojunction.

In the region of an $EC_b$/SE-AdSIC heterojunction mobile cations overcome a sequence of 27 potential barriers. Assume that the misfit stresses on the $EC_b$/$\{X^j\}$ coherent heterojunction disappear at distances $\approx 2$ lattice parameters ($\approx 1$ nm for $\alpha$-AgI AdSIC). A major drop of $\eta$ (Fig.3) occurs at this distance. The envelope of heights of barriers on the $EC_b$/$\{X^j\}$ is the exponent that minimizes the number of model parameters

$$\eta_{j,j+1} = a_1 + b_1 \cdot \exp(-j/c), \quad j = 1,2,..\ 27, \qquad (1)$$

where $a_1 = 0.1$ eV and $b_1 = 0.4$ eV. The "$a$" parameter determines the barrier heights $\eta_{j,j+1}$ in the AdSIC volume and the $a + b$ sum defines the height of the largest barrier $\eta_{1,2}$ in the sequence of planes $\{X^j\}$ ($j = 1,2…28$). The value $c = 7$ provides the envelope value equal to the barrier heights in the AdSIC volume. The boundary electrode $EC_b$ is denoted as $X^0$. The $X^{28}$ plane of a rigid AdSIC sub-lattice situated in place where the height of potential barrier $\eta_{27,28}$ reaches the value equal to that in the volume $\approx 0.1$ eV. The boundary plane $X^{28}$ can also be considered as an ideally kinetically reversible electrode $EC_r$ on which direct and reverse electrochemical reactions of mobile cations occur at infinitesimal overvoltage (electrode impedance equals zero). The external influence on the heterojunction was exerted by a current generator operating in the galvanostatic mode.

2. A few strained monolayers of solid electrolyte (SE) allocated between two AdSICs forming coherent interfaces with SE. A crystallographically narrow coherent heterojunctions of the AdSIC/SE/AdSIC type models processes in an non-uniform potential relief arising due to an exponential drop of elastic stresses in SE. The strain and strain-barrier height $\eta_{j,j+1}$ correspond to these stress. The edge barriers $\eta_{1,2} = 0.2$ eV and $\eta_{27,28} = 0.2$ eV belong to AdSIC with the potential barriers 0.1 eV in the volume. We consider the edge interfaces between AdSICs and SE as coherent and ideally reversible electrodes for mobile ions. Such objects are hard to study by impedance spectroscopy which gives coarse averaged data (by volume and time) for interconnected ion-transport and dielectric-polarization parallel-serial processes (collective phenomenon).

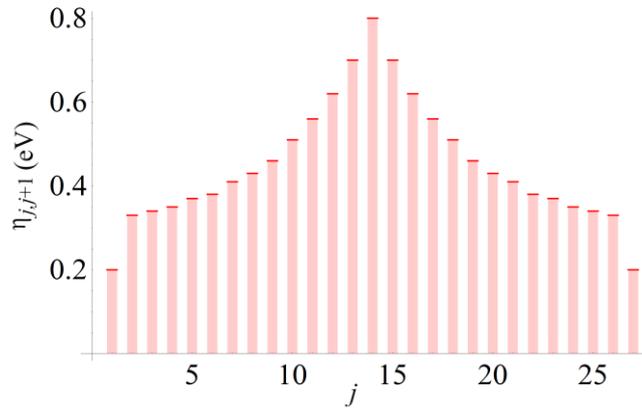

**Figure 4**: The profile of a potential relief of model AdSIC/SE/AdSIC nanostructure, i.e. a few strained monolayers of a solid electrolyte (SE) allocated between two AdSICs forming coherent interfaces with SE.

External influence on the AdSIC/SE/AdSIC nanostructure is specified by ac current generator with the amplitude of current density $I_0 = const$ ($\omega$). Radian frequency $\omega$ was in the range of impedance spectroscopy. Computer experiments were performed in the "Wolfram Mathematica" package.

### 4. Maxwell displacement current on a potential barrier

The structure-dynamic approach assumes that a uniform effective electric field along the $x$ axis operates in a layer between $X^j$ and $X^{j+1}$ planes [20]. The electric field strength $F_{j+1,j}$ satisfies the Gauss law

$$F_{j+1,j} = \frac{-en_0}{\varepsilon_0 \varepsilon_{j,j+1}} \sum_{k=j+1}^{M} y_k , \qquad (2)$$

where $y_k(t) \equiv (n_k - n_0)/n_0$ is a relative change in the concentration of ions (cations) of mobile kind in the minimum of $k$ of a potential relief, $n_k$ ($n_0$) is a non-equilibrium (equilibrium in the absence of electric field) concentration of mobile cations in the minimum $k$, $e$ is an absolute value of an electron charge, $\varepsilon_{j,j+1}$ is an dielectric susceptibility, $\varepsilon_0$ is an electric constant. The total charge $en_0 \Sigma y_k$ on an ideally polarizable heterojunction is created by a current generator. The sign "minus" in eq. (2) shows that at $y_k < 0$ (deficit of cations on $\{X^k\}$ planes) the $F_{j,j+1}$ vector and the positive direction along the $x$ axis coincide. The vector of electric induction in the layer between $X^j$ and $X^{j+1}$ planes is given by the equation

$$D_{j+1,j} = \varepsilon_0 \varepsilon_{j,j+1} F_{j+1,j} . \qquad (3)$$

The local Maxwell displacement current $I_{Dj,j+1}$ on the barrier $\eta_{j,j+1}$ can be determined through the time derivative of the $x$- component of the $D_{j+1,j}$ vector

$$I_{D_{j,j+1}}(t) \equiv \frac{dD_{j+1,j}}{dt} \qquad (4)$$

Computer experiments in terms of the structure-dynamic approach showed that the displacement current $I_{Dj,j+1}$ satisfies, with an accuracy of more than $10^{-5}$ %, the condition of a circuit closed on a current generator

$$I_{j,j+1}(t) + I_{D_{j,j+1}}(t) = I_M(t), \qquad (5)$$

i.e. the sum of an ionic current $I_{j,j+1}$ and a displacement current $I_{Dj,j+1}$ on each barrier $\eta_{j,j+1}$ equals an external current $I_M(t)$ which is created by a generator on the heterojunction boundaries.

Equation (5) can be proved analytically. This means that the structure-dynamic approach should include conditions for equation (5) to be valid for any potential barrier $\eta_{j,j+1}$ on an ideally polarizable heterojunction. Indeed, the basic kinetic equation (equation (7) in [22]) can be expressed as an ionic current $en_0 dy/dt$ through the $X^j$ plane ($x_j$ potential minimum) and ionic currents over adjacent potential barriers $I_{j-1,j}$ and $I_{j,j+1}$

$$en_0 \frac{dy_j}{dt} = I_{j-1,j} - I_{j,j+1} , \qquad (6)$$

where $I_{j-1,j} \equiv I_{j \rightarrow j-1} + I_{j \leftarrow j-1}$ and $I_{j,j+1} \equiv I_{j \rightarrow j+1} + I_{j \leftarrow j+1}$ are the resulting densities of cation currents over the barriers $\eta_{j,j-1}$ and $\eta_{j,j+1}$, expressed by the current densities in the $+x$ direction ($I_{j \rightarrow j+1}$ и $I_{j \rightarrow j-1}$) and in the $-x$ direction ($I_{j \leftarrow j+1}$ и $I_{j \leftarrow j-1}$). The sign "minus" in (6) means that the cation concentration (positive charge) on the $X^j$ plane decreases if the resulting current vector $I_{j,j+1}$ coincides with the positive direction of $x$, i.e. $+x$ axis. Taking eqs. (2), (3) and (4) into account, the expression for $I_{Dj,j+1}(t)$ can be written as

$$I_{D_{j,j+1}} = -en_0 \sum_{k=j+1}^{M} \frac{dy_k}{dt} . \qquad (7)$$

Now consider an $EC_b/AdSIC/X^M$ nanostructure with a blocking $EC_b/AdSIC$ heterojunction ($x_0$ in Fig.1) and with right edge $x_M$. The movement of cations in the $+x$ direction occurs at $I_M(t) > 0$. For the nanostructure $EC_b/AdSIC/X^M$ the integral with the upper variable limit $\int_0^t I_M(t)\,dt$ (i.e. charge accumulated on the heterojunction) can be exactly substituted by the summation of charge on all potential minima

$$\int_0^t I_M(t)dt = -en_0 \sum_{k=1}^{M} y_k \qquad (8)$$

Differentiation of (8) with respect to the upper limit gives

$$I(t) = -en_0 \sum_{k=1}^{M} \frac{dy_k}{dt} = -en_0 (\sum_{k=1}^{j} \frac{dy_k}{dt} + \sum_{k=j+1}^{M} \frac{dy_k}{dt}) \qquad (9)$$

After summation, eq. (9) takes on the form

$$I(t) = I_{Dj,j+1} - (I_{0,1} - I_{j,j+1}) \; . \qquad (10)$$

By definition, the ionic current of an ideally polarizable electrode $EC_b$ is $I_{0,1} = 0$. Hence, equation(10) transforms into (5), i.e. it is analytically proved that the sum of ionic and displacement currents is equal to the current of an external generator for any potential barriers $\eta_{j,j-1}$ in any moment of time.

## 5. Computation data and their discussion

### 5.1. Modeling of processes in the galvanostatic mode

Figure 5 presents the calculated data for the time dependence of capacitance, Maxwell displacement currents $I_{Dj,j+1}(t)$ and ionic conductivity currents $I_{j,j+1}(t)$ corresponding to the charging of an $EC_b/\{X^j\}$ ($j = 1,2…28$) in the galvanostatic mode: $I(t) = const$ at $t > 0$ and $I(t) = 0$ at $t \leq 0$. The computations were performed assuming that the charge accumulated on the heterojunction is small (steric effects are absent)

$$\int_0^t I(t)dt \ll en_0 \qquad (11)$$

The calculations demonstrate (Fig.5) that the conditions of a closed circuit (5) hold for six highest barriers $\eta_{j,j+1}$ of the $EC_b/\{X^j\}$heterojunction. It is seen in the figure that the displacement currents $I_{Dj,j+1}(t)$ dominate over the ionic conductivity currents $I_{j,j+1}(t)$ at small $t$. The condition $I_{Dj,j+1} = I_{j,j+1}$ is reached at the characteristic time $t^*_{j,j+1}$ depending on the barrier height $\eta_{j,j+1}$. At $t \gg \max\ t^*$ the space charge on the heterojunction is characterized by a stationary non-equilibrium distribution with $y_j/y_{j+1} \approx const$. Capacitance $C$, determined by the ratio of the accumulated charge to the voltage on the heterojunction, depends on $\eta_{j,j+1}$ due to the proportionality $\varepsilon_{j,j+1} \propto 1/\eta_{j,j+1}$[10] and reaches its maximum at $t_{max} > t^*_{1,2}$ where the $t_{max}$ value depends on the external current $I(t)$ because of the steric effect.

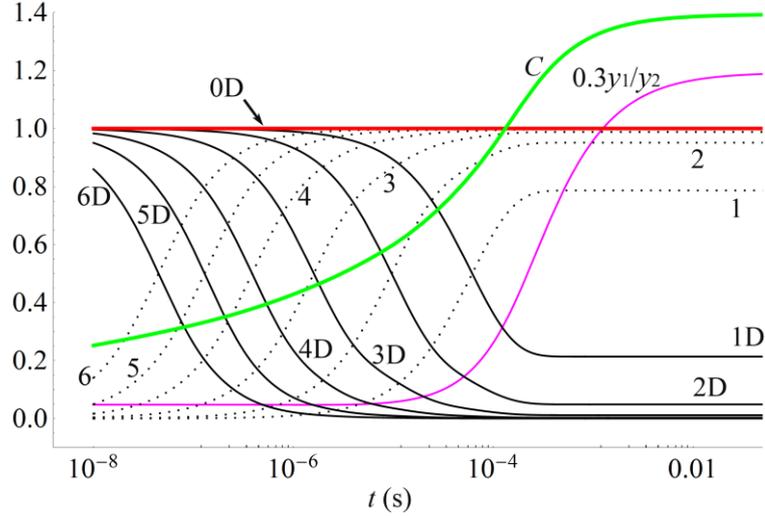

**Figure 5**: Time dependence of the capacitance $C$ (F m$^{-2}$) of ionic space charge, ratio $0.3y_1/y_2$ (coefficient 0.3 for clearness), normalized Maxwell's displacement currents ($I_{Dj,j+1}/I(t)$, solid lines) and normalized cation currents ($I_{j,j+1}/I(t)$, dotted lines) in the region of the EC$_b$/{$X^j$} ($j = 1,2…28$) blocking heterojunction charging in the galvanostatic mode. The marker numbers from 1 to 6 correspond to the ionic current $I_{j,j+1}/I(t)$ over barrier $\eta_{j,j+1}$ ($j = 1,2,…6$), the marker D corresponds to normalized Maxwell displacement current $I_{Dj,j+1}/I(t)$ through barrier $\eta_{j,j+1}$ ($j = 1,2,…6$), the marker 0D corresponds to normalized Maxwell displacement current between the EC$_b$ and minimum potential relief at $x_1$.

The data obtained suggest that the heights of potential barriers $\eta_{j,j+1}$ in the region of a heterojunction should be small in devices with high frequency-capacitance characteristics. Small $\eta_{j,j+1}$ values can be achieved by the choice of contacting materials, i.e. EC$_b$/AdSIC, and creation of a coherent (semi-coherent) heteroboundary [8,27,31].

The processes modeled in this work, in which the displacement current $I_{Dj,j+1}$ and ionic conductivity currents $I_{j,j+1}$ are interconnected, are determined by mutually consistent behavior of "hidden" variables - the concentrations of mobile cations $n_j(t)$ in the minima of a potential relief $x_j$. Static steric effects [14,15] as well as dynamic ones can be modeled in the frame of the structure-dynamic approach. Static effects become appreciable when sufficiently large part of cations leave the potential minimum with $j = 1$, nearest to the EC$_b$. In this case the cations begin to leave the minimum with $j = 2$ also. Steric effects become pronounced when a space charge capacitance $C$ passes through a maximum at the condition

$$\int_0^t I(t)\mathrm{d}t \sim en_0. \qquad (12)$$

However, for obtaining correct data on steric effects in strongly non-equilibrium dynamic conditions it is necessary to determine the errors connected with the application of the detailed balance principle. In this work, in modeling of the dynamic processes the condition $\int_0^t I(t)\,\mathrm{d}t \ll en_0$ (11) is fulfilled. This permits the principle of detailed balance to be correctly used.

Figure 6 shows that at $t \approx 0$ the growth of relative changes in the cation concentrations $y_j(t) \equiv (n_j - n_0)/n_0$ on the $X^j$ planes with small $j$ indices (e.g. $j =1, 2, 3$ and 4, large $\eta_{j,j+1}$) occur without a lag relatively to changes of $y_j(t)$ with large indices (e.g. $j = 8$ and 9, small $\eta_{j,j+1}$) during a heterojunction charging by a current step (galvanostatic mode). This means that the long-range electric field created by ionic charge of the $X^{28}$ edge plane determines EDL-relaxation at $t \approx 0$.

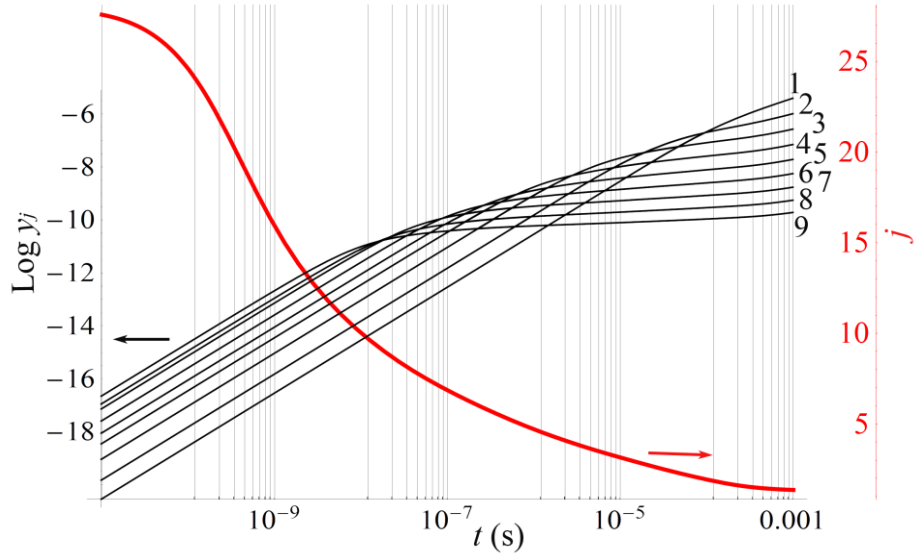

**Figure 6**: The time dependence of relative concentrations $y_j$ ($j = 1, 2, 3, 4, 5, 7, 8$ and 9) in the double logarithmic scale (left ordinate) for galvanostatic charging of the $EC_b/\{X^j\}$ ($j = 1,2…28$) ideally polarized heterojunction ($T = 300$ K, $n_0 = 10^{18}$ m$^{-2}$). The graph number corresponds to the index $j$ of a minimum potential relief. Right ordinate shows the calculated center $\lambda(t)$ of mass of mobile cations in EDL (in the units of $x/\Delta$, i.e. $j$ index).

The polarization length ($\lambda$) which is often used to characterize the layer thickness of an ionic space charge in electrochemistry can be connected with the position of the center of mass of mobile cations in EDL (Fig.6). The conditions $y_{j+1}(t) = y_j(t)$, i.e. points of crossing of adjacent curves in Fig.6, give rough estimates of time dependent $\lambda(t)$. We propose to determine $\lambda(t)$ from the formula of the mass center position of a space cation charge calculated with respect to the $X^0$ planes, i.e. the $EC_b$ surface

$$\lambda(t) \equiv \frac{\Delta \sum_j j y_j(t)}{\sum_j y_j(t)} \;. \tag{13}$$

Replacing "$t$" by "$2\pi/\omega$" in eq. (13), we obtain a frequency-dependent polarization length $\Lambda(\omega)$
$$\Lambda(\omega) \equiv \lambda(2\pi/\omega) \;. \tag{14}$$

The polarization length $\lambda(t)$ tends to the length of charge screening $\lambda_\infty$ at $t \gg t^*_{1,2}$ (characteristic time for the highest barrier). Calculations of free relaxation of a non-equilibrium space charge on the $EC_b/AdSIC$ (i.e. under the condition $e\Sigma n_i = const$) show [22,25] that the capacitance of a heterojunction is an increasing function of time, like in the case of galvanostatic charging. This capacitance behavior agrees with a great number of experimental data reported in the literature (e.g., [10,32,33]). On the other hand, the description of space charge relaxation in terms of the structure-dynamic approach disagrees with the interpretation in [32,33] based on the model of

EDL adsorption relaxation and the concept of minor charge carriers. According to that model, large adsorption capacitance $C_{ad} = const$ ($C_{ad} \gg C = const$) and Warburg impedance are introduced into equivalent electrical circuits of $EC_b$/AdSIC heterojunctions to fit the calculated and experimental data on the blocking heterojunction response to low-frequency external influence.

### *5.2. Jonsher's "universal" dynamic response in nanoionics*

Introduction of the notion of the local Maxwell displacement currents [22,24] into the nanoionics concept [26] permits a new viewpoint on the dynamic phenomena and effects which nature and mechanisms remain poorly understood. One of still unsolved problem of solid state ionics is the "universal" dynamic response.

Since the review by A.K. Jonsher [34], the frequency dependence of a linear electric response attracts much attention and is considered until now as fundamental [35-41]. A new field of investigations appeared. Its subject is the dynamics of ion transport in solids [2,4,35]. The discovered power law of the real part of frequency dependent conductivity $\text{Re}\sigma^*(\omega) \propto \omega^n$ ($n < 1$) is usually used in standard interpretations in materials science and solid state ionics [42-53]. However, understanding of the dynamics of ion transport in disordered materials, i.e. FIT on nanoscale, remains an unsolved problem. This can be explained by the fact that existent experimental data cannot be interpreted unambiguously. A widely used and most informative method of impedance spectroscopy gives experimental data which are roughly volume and time averaged. At averaging, a large amount of information (which reflects interconnected local processes of ion transport and dielectric polarization [54,55]) is lost.

Figure 7 presents the results of computer modeling of the ac current generator influence (amplitude current density $I_0 = const$) on the formation of space charge, which occurs in the region of a non-uniform potential relief in the AdSIC/SE/AdSIC nanostructure of $L$ width.

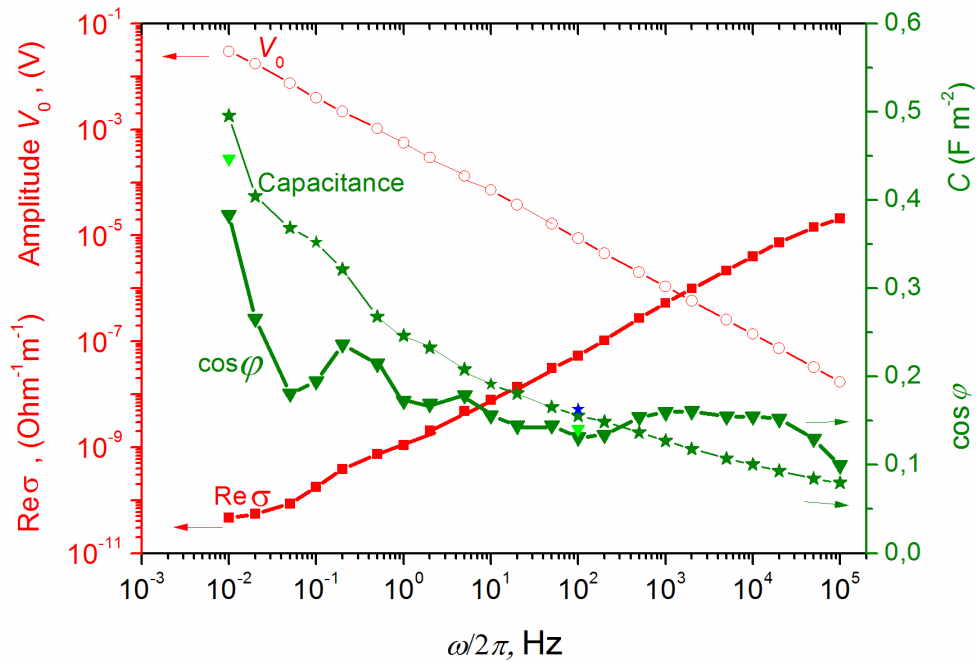

**Figure 7**: Frequency dependence of Log $\text{Re}\sigma^*$ (in $Ohm^{-1} m^{-1}$), Log $V_0$ (in V), ionic space charge capacitance $C$ (in $F m^{-2}$) and $\cos\varphi$ where $\varphi$ is the phase shift between current density in AdSIC/SE/AdSIC and voltage on the edges of the nanostructure.

The value of $\text{Re}\sigma^*(\omega)$ for the AdSIC/SE/AdSIC varies with the frequency $\omega$ approximately by the power law

$$\text{Re}\sigma^*(\omega) \equiv (L/V_0)I_0\cos\varphi \propto \omega^n \qquad (15)$$

where $n < 1$, i.e. demonstrates the 'universal' dynamic response in a wide frequency range.

Analysis of the considered phenomenon (power law) in terms of hidden variables $n_j$, non-equilibrium concentrations of mobile ions on the crystallographic planes $X^j$ in the region of a non-uniform potential relief shows:

(i) The charge density on $X^j$ planes, i.e. $\Delta q_j = e\,(n_j - n_0)$, varies as integral under the influence of an external ac current

$$\Delta q_j \propto I_0 \int \sin(\omega t)\,dt. \qquad (16)$$

The integration with respect to $t$ gives $\Delta q_j \propto 1/\omega$;

(ii) The charges $\Delta q_j$ determine the voltage between the $X^j$ and $X^{j+1}$ planes and, as a result, the amplitude of $V_0$ total ac voltage varies approximately as $1/\omega$;

(iii) The amplitude $V_0(\omega)$ gives a dominant contribution to $\text{Re}\sigma^*(\omega)$;

(iv) The phase shift $\varphi$ between the current generator $I(t) = I_0 e^{i\omega t}$ and voltage $V_{27,1}(t) = V_0 e^{i(\omega t + \varphi)}$ on the model nanostructure is determined by averaging a set of local parallel-serial processes of ionic transport and dielectric polarization, i.e. interconnection of local conductivity currents and the Maxwell displacement currents.

Figure 8a, 8b and 8c present the distribution of the Maxwell displacement currents for a non-uniform potential relief in the region of AdSIC/SE/AdSIC nanostructure for three frequencies of current generator ($\omega_1 \gg \omega_2 \gg \omega_3$) at the condition $I(t) = I_{j,j+1}(t) + I_{Dj,j+1}(t)$.

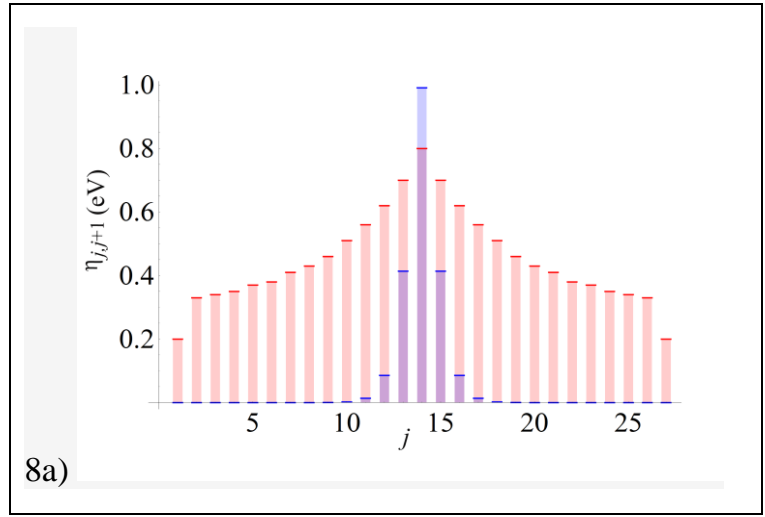

8a)

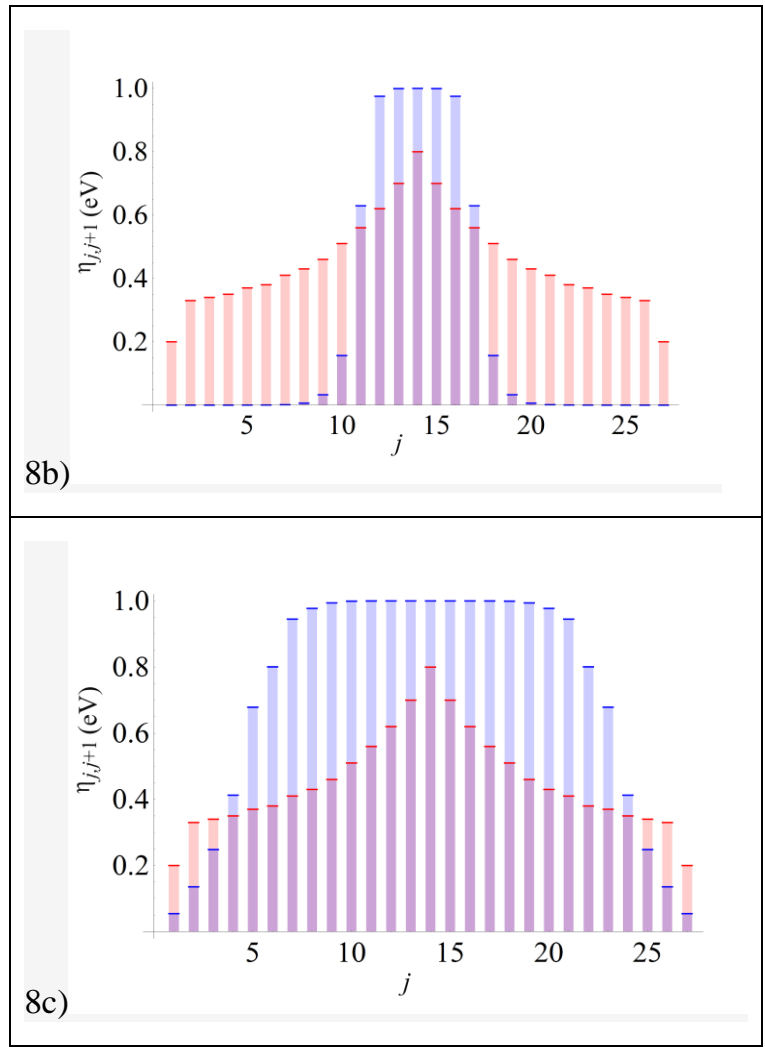

8b)

8c)

**Figure 8**: The ratio (violet color) of amplitudes of the Maxwell displacement currents $I_{Dj,j+1}$ on potential barriers to amplitude $I_0$ of current of current generator for the AdSIC ($j =1$)/SE ($j = 2 \div 26$)/AdSIC ($j =27$) nanostructure with a non-uniform potential relief (pink color): a) a current generator with frequency $\omega/2\pi = 0.01$ Hz, $\cos\varphi = 0.383$; b) $\omega/2\pi = 10$ Hz, $\cos\varphi = 0.156$; c) $\omega/2\pi = 50$ kHz, $\cos\varphi = 0.129$.

Comparison of Figures 8a, 8b and 8c with the data in Fig. 7 shows that the cosine of $\varphi$ weakly depends on $\omega$ in the cases when the local Maxwell displacement currents and ionic conductivity currents are of the same order in a set of potential barriers in the region of a non-uniform potential relief. The power law holds in a wide frequency range if the difference of potential barrier heights is rather large whereas the distribution of barriers in height (not spatial disorder in potential relief) is rather smooth.

The influence of spatial disorder of barriers on the behavior of $\text{Re}\sigma^*(\omega)$ is not strong. It will be considered in other work.

## 6. Conclusions

A simple theoretical approach to the nanoscale dynamic modeling of ion-transport in a non-uniform potential relief is generalized by including a new notion - the Maxwell displacement current on potential barrier. "Simple" means that the approach is based on the clear physical approximations, easily simulates a wide range of dynamic and static ion-transport properties and collective phenomena, while calculated results are in good agreement (without large number of fitting parameters) with the known literature experimental data. The work proves (by analytically and in compute experiment) that the Maxwell displacement currents $I_{Dj,j+1}(t)$ on any potential barriers together with ionic current $I_{j,j+1}$ over the same barrier satisfy the condition of an electric circuit closed on a current generator with current density $I(t)$, i.e. for any barrier $\eta_{j,j+1}$ the equality $I(t) = I_{j,j+1}(t) + I_{Dj,j+1}(t)$ holds. In scientific literature, Jonsher's "universal" dynamic response is one of fundamental problems of solid state ionics and materials science. The application of the structure-dynamic approach for simulation of coupled ion-transport and dielectric-polarization processes (collective phenomena) in the region of non-uniform potential relief shows clearly the factors and conditions which define the origin of power law, i.e. $\mathrm{Re}\sigma^*(\omega) \propto \omega^n$ ($n < 1$), in the behavior of the real part of frequency dependent complex conductivity.

**References**

1. H. Mehrer, Diffusion in Solids, Springer, Berlin Heidelberg New York, 2007.

2. J. C. Dyre, P. Maass, B. Roling, and D. L. Sidebottom, "Fundamental questions relating to ion conduction in disordered solids," Reports on Progress in Physics, vol. 72, no. 4, pp. 046501-1–46501-15, 2009.

3. D. L. Sidebottom, "Colloquium: Understanding ion motion in disordered solids from impedance spectroscopy scaling," Reviews of Modern Physics, vol. 81, no. 3, pp. 999–1014, 2009.

4. J. R. Macdonald, "Addendum to "Fundamental questions relating to ion conduction in disordered solids"," Journal of Applied Physics, vol. 107, no. 10, pp. 101101-1–101101-9, 2010.

5. L. Bindi, M. Evain, A. Pradel, S. Albert, M. Ribes, and S. Menchetti, "Fast ion conduction character and ionic phase-transitions in disordered crystals: the complex case of the minerals of the pearceite–polybasite group," Physics and Chemistry Minerals, vol. 33, no. 33, pp. 677–690, 2006.

6. D. D. Fong, G. B. Stephenson, S. K. Streiffer, J. A. Eastman, O. Auciello, P. H. Fuoss, and C. Thompson, "Ferroelectricity in ultrathin perovskite films," Science, vol. 304, no. 5677, pp. 1650–1653, 2004.

7. A. Boulant, J. Emery, A. Jounanneaux, J.-Y. Buzare, and J.-F. Bardeau, "From micro- to nanostructured fast ionic conductor Li0.30La0.56□ 0.13Tio3: Size effects on NMR properties," The Journal of Physical Chemistry C, vol. 115, no. 31, pp. 15575−15585, 2011.

8. A.L. Despotuli, A.V. Andreeva, and B. Rambabu, "Nanoionics of advanced superionic conductors," Ionics, vol. 11, no. 3-4, pp. 306–314, 2005.

9. A.L. Despotuli, and A.V. Andreeva, "A short review on deep-sub-voltage nanoelectronics and related technologies," International Journal of Nanoscience, vol. **8**, no. 4-5, pp. 389–402, 2009.

10. A.L. Despotuli, and A.V. Andreeva, "Nanoionics: new materials and supercapacitors", Nanotechnologies in Russia, vol. 5, no. 7–8, pp. 506–520, 2010.

11. S.H. Liu, "Lattice gas model for the metal - electrolyte interface," Surface Science, vol. 101, no. 1−3, pp. 49−56, 1980.

12. J.R. Macdonald, "Layered lattice gas model for the metal - electrode interface," Surface Science, vol. 116, no. 1, pp. 135−147, 1982.

13. Macdonald J.R., and Liu S.H. "An iterated three-layer model of the double layer with permanent dipoles", Surface Science, vol. 125, no. 3, pp. 653–678, 1983.

14. M.S. Kilic, M.Z. Bazant, and A. Ajdary, "Steric effects in the dynamics of electrolytes at large applied voltages: I. Double-layer charging," Physical Review E, vol. 75, no. 2, pp. 021502-1– 021502-16, 2007.

15. M.Z. Bazant, B.D. Storey, and A.A. Kornyshev, "Double layer in ionic liquids: Overscreening versus crowding," Physical Review Letters, vol. 106, no. 4, pp. 046102-1– 046102-42011, 2011.

16. S.A. Kislenko, R.N. Amirov, and I.S. Samoylov, "Molecular dynamic simulation of the electrical double layer in ionic liquids", Journal of Physics: Conference Series, vol. 418, pp. 012021-1– 012021-8, 2013.

17. I. Borukhov, D. Andelman, and H. Orland, "Steric effects in electrolytes: A modified Poisson-Boltzmann equation," Physical Review Letters, vol. 79, no. 3, pp. 435–438, 1997.

18. M.B. Singh, and R. Kant "Theory of electric layer dynamics at blocking electrode ," arXiv:1103.0681v1 [cond-mat.mtrl-sci], 2011.

19. M. Soestbergen, P.M. Biesheuvel, M.Z. Bazant, "Diffuse-charge effects on the transient response of electrochemical cells," Physical Review E, vol. 81, no. 2, pp. 021503-1–021503-13, 2010.

20. A.L. Despotuli, and A.V. Andreeva, "Model, method and formalism of new approach to ion transport processes description for the solid electrolyte/electronic conductor blocking heterojunctions," Nano and Microsystem Technique (Russian), no, 9, pp. 16–21, 2012. http://www.nanometer.ru/2013/08/22/nanoionika_333471.html

21. A.L. Despotuli, and A.V. Andreeva, "Computer modeling on sub-nanometer scale the ion-transport characteristics of the solid electrolyte/electronic conductor blocking heterojunctions," Nano and Microsystem Technique (Russian), no. 11, pp. 15–23, 2012. http://www.nanometer.ru/2013/08/22/nanoionika_333471.html

22. A.L. Despotuli, and A.V. Andreeva, "Maxwell displacement current in nanoionics and intrinsic ion-transport properties of model nanostructures," Nano and Microsystem Technique (Russian), no. 8, pp. 2–9, 2013. http://www.nanometer.ru/2013/08/22/nanoionika_333471.html

23. A.L. Despotuli, and A.V. Andreeva, "Modeling on sub-nanometer scale of fast ionic transport processes on blocking heterojunctions of solid electrolyte/ electronic conductor," Electronic Journal: Phase transformations, ordering states and new materials (Russian), no. 2, pp. 7–12, 2013. http://ptosnm.ru/ru/issue/2013/2/83/publication/776

24. A.L. Despotuli, and A.V. Andreeva, "Maxwell displacement current in nanoionics. Modeling of processes of space charge relaxation on blocking heterojunctions of solid electrolyte/ electronic conductor," Electronic Journal: Phase transformations, ordering states and new materials (Russian), no. 10, pp. 26–40, 2013. http://ptosnm.ru/ru/issue/2013/10/91/publication/856

25. A.L. Despotuli, and A.V. Andreeva, "Structure-dynamic approach in nanoionics. Modeling of ion transport on blocking electrode," arXiv:1311.3480 [cond-mat.mtrl-sci]

26. A.L. Despotuli, and V.I. Nikolaichik, "A step towards nanoionics," Solid State Ionics, vol. 60, no. 4, pp. 275–278, 1993.

27. A.V. Andreeva, and A.L. Despotuli, "Interface design in nanosystems of advanced superionic conductors," Ionics, vol. 11, no. 1–2, pp. 152–160, 2005.

28. W. H. Flygare, and R. A. Huggins, "Theory of ionic transport in crystallographic tunnels," Journal of Physics and Chemistry of Solids, vol. 34, no. 7, pp. 1199-1204, 1973.

29. A.J. Ardell, "Gradient energy, interfacial energy and interface width," Scripta Materialia, vol. 66, no. 7, pp. 423–426, 2012.

30. K.-R. Lee, K. Ahn, Y.-C. Chung, J.-h. Lee, H.-I. Yoo, "Lattice distortion effect on electrical properties of GDC thin films: experimental evidence and computational simulation," Solid State Ionics, vol. 229, no. 12, pp. 45–53, 2012.

31. A.L. Despotuli, and A.V. Andreeva,,"Creation of new types of thin-film solid electrolyte supercapacitors for microsystems technology and micro (nano)electronics," Microsystem Technique (Russian), no. 11, pp. 2–10, 2003.

32. E. A. Ukshe, and N.G. Bukun, "Development of the model of adsorptive double-layer relaxation in superionic conductors," Soviet Electrochemistry, vol. 26, no.11, pp. 1221–1229, 1990.

33. N.G. Bukun, and A.E. Ukshe, "Impedance of solid electrolyte systems," Russian Journal of Electrochemistry, vol. 45, no. 1, pp. 11–24, 2009.

34. A.K. Jonscher, "The 'universal' dielectric response," Nature, vol. 267, no. 6, pp. 673–679, 1977.

35. K. Funke, "Solid State Ionics: from Michael Faraday to green energy—the European dimension," Science and Technology of Advanced Materials, vol. 14, no. 4, pp. 043502–, 2013.

36. Z. Wojnarowska, A. Swiety-Pospiech, K. Grzybowska, L. Hawelek, M. Paluch, and K. L. Ngai, "Fundamentals of ionic conductivity relaxation gained from study of procaine hydrochloride and procainamide hydrochloride at ambient and elevated pressure," Journal of Chemical Physics, vol. 136, no. 16, pp. 164507–, 2012.

37. I. I. Popov, R. R. Nigmatullin, A. A. Khamzin, and I. V. Lounev, "Conductivity in disordered structures: Verification of the generalized Jonscher's law on experimental data," Journal of Applied Physics, vol. 112, no. 9, pp. 094107–, 2012.

38. J. Sibik, E.Y. Shalaev, and J. A. Zeitler, "Glassy dynamics of sorbitol solutions at terahertz frequencies," Physical Chemistry Chemical Physics, vol. 15, no. 5, pp. 11931–11942, 2013.

39. J.C. Dyre, "Aging of CKN: Modulus versus conductivity analysis," Physical Review Letters, vol. 110, no. 24, pp. 245901–, 2013.

40. Y.-H. Rim, M. Kim, J. E. Kim, and Y. S, Yang, "Ionic transport in mixed-alkali glasses: hop through the distinctly different conduction pathways of low dimensionality," New Journal of Physics, vol. 15, no. 2, pp. 023005–, 2013.

41. C. Mattner, B. Roling, and A. Heuer, "The frequency-dependence of nonlinear conductivity in disordered systems: an analytically solvable model," arXiv:1302.3258v1[cond-mat.dis-nn].

42. R.D. Banhatti, , D. Laughman, L. Badr, K. and Funke, "Nearly constant loss effect in sodium borate and silver meta-phosphate glasses: New insights," Solid State Ionics, vol. 192, no.1, pp. 70–75, 2011.

43. S. Ke, P. Lin, H. Huang, H. Fan, and X. Zeng, "Mean-field approach to dielectric relaxation in giant dielectric constant perovskite ceramic," Journal of Ceramics, vol. 2013, Article ID 795827, 7 pages, 2013.

44. T. V. Kumar, A. S. Chary, S. Bhardwaj, A. M. Awasthi, and S. N. Reddy, "Dielectric relaxation, ionic conduction and complex impedance studies on $NaNO_3$ fast ion conductor," International Journal of Materials Science and Applications, vol. 2, no. 6, pp. 173–178, 2013.

45. B. N. Parida, P. R. Das, R. Padhee, and R. N. P. Choudhary, "Structural, dielectric and electrical properties of $Li_2Pb_2La_2W_2Ti_4Nb_4O_{30}$ ceramic," Bulletin of Materials Science, vol. 36, no. 5, pp. 883–892, 2013.

46. M. M. A. Kader, F. El-Kabbany, H. M. Naguib, and W. M. Gamal, "Charge transport mechanism and low temperature phase transitions in $KIO_3$," Journal of Physics: Conference Series, vol. 423, pp. 012036– , 2013.

47. A. Kriaa, K. B. Saad, and A. H. Hamzaoui, "Synthesis and characterization of cancrinite-type zeolite, and its ionic conductivity study by AC impedance analysis," Russian Journal of Physical Chemistry A, vol. 86, no. 3, pp. 2024–2032, 2012.

48. A. Šantić, and A. Moguš-Milanković, "Charge carrier dynamics in materials with disordered structures: a case study of iron phosphate glasses," Croatica Chemica Acta, vol. 85, no. 3, pp. 245–254, 2012.

49. M. Chaari, and A.Matoussi, "Electrical conduction and dielectric studies of ZnO pellets," Physica B, vol. 407, no. 9, pp. 3441–3447, 2012.

50. C. Peng-Fei, L. Sheng-Tao, L. Jian-Ying, D. Can, and Y. Yan, "Physical meaning of conductivity spectra for ZnO ceramics," Chinese Physics B, vol. 21, no. 9, pp. 097201–, 2012.

51. N.I.Sorokin, and B.P,Sobolev, " Frequency Response of the Low-Temperature Ionic Conductivity of Singe Crystals $R_{1-y} M_y F_{3-y}$ (R= La-Er; M= Ca, Sc, Ba, Cd)," Physics of the Solid State, vol. 50, no. 3, pp. 402–407, 2008.

52. D. Regonini, V. Adamaki, C.R. Bowen,, S.R. Pennock, J. Taylor, and A.C.E. Dent, "AC electrical properties of $TiO_2$ and Magnéli phases, $Ti_nO_{2n-1}$," Solid State Ionics, vol. 229, pp. 38–44, 2012.

53. D. P. Singh, K. Shahi, K. K. Kar, "Scaling behavior and nearly constant loss effect in AgI–$LiPO_3$ composite glasses," Solid State Ionics, vol. 231, pp. 102–108, 2013.

54. C. R. Bowen, and D. P. Almond, "Modelling the 'universal' dielectric response in heterogeneous materials using microstructural electrical networks," Materials Science and Technology, vol. 22, no. 6, pp. 719–724, 2006.

55. F. Henn, S. Devautour-Vinot, J.-G. Giuntini, J. Bisquert, G. Garcia-Belmonte, C. Platon, E. Varsamis, and E.l. Kamitsos, "Analysis of AC permittivity response measured in an ionic glass: A comparison between iso and non-iso thermal conditions," IEEE Transaction on Dielectrics and Electrical Insulation, vol. 17, no. 4, pp. 1164–1171, 2010.